# The case for extant life on Mars: A Technical Reanalysis of the Viking Carbon Assimilation–Pyrolytic Release Experiment

Jan Spacek, Clay A. Abraham, and Steven A. Benner*
Foundation for Applied Molecular Evolution, 13709 Progress Drive, Alachua, FL 32615

jspacek@ffame.org
cabraham@ffame.org
*sbenner@ffame.org, corresponding author

## Abstract

The 1976 Viking Carbon Assimilation–Pyrolytic Release (CAPyR) experiment tested whether carbon supplied as $^{14}CO_2$ and $^{14}CO$ would, over a 120-hour incubation at low temperatures, enter Martian soil in a form retained during flushing at 120 °C but released upon pyrolysis at 635 °C and captured in an organic vapor trap (Chromosorb-P/CuO). Untrapped radioactivity, presumed to be $^{14}CO_2$ and/or $^{14}CO$, was measured as Peak 1; trapped material was subsequently oxidized by CuO and measured as Peak 2. According to the design, net Peak 2 was to be interpreted as representing reduced organic material synthesized from $^{14}CO_2$ and $^{14}CO$, presumptively by Martian biology. At Chryse and Utopia Planitia, seven of nine usable incubations gave significant net Peak 2 signals, including the dark U1 run. A documented VL2 gas leak introduces unresolved uncertainty into the U1-U3 results. In C1, the signal corresponded to ~10 pmol of carbon if derived from CO or ~30 pmol if derived from $CO_2$, approximately $10^3$ *Escherichia coli*-equivalents of carbon. Subsequent recognition that perchlorate salts may have been present in Viking soils may require reconsideration of how the CAPyR data are interpreted. Oxidation of fixed $^{14}C$-labeled organic carbon during pyrolysis may have shifted, or "repartitioned," radiolabel away from Peak 2 and into Peak 1. This perchlorate-specific repartitioning was not considered in previous CAPyR analyses. Hubbard's (1979) mineral-substrate experiments generated abiotic Peak 2 signals under Viking-relevant conditions without reproducing the complete Viking dataset. These data are examined in detail. No demonstrated abiotic mechanism has reproduced all CAPyR observations under Mars-relevant conditions, while the observations remain consistent with a biological carbon-assimilation system whose relevant activity survives C5's 120 °C/90 °C pretreatment but is strongly impaired after C2's 175 °C pretreatment. Together with the Labeled Release experiment, in which an oxidative process released radiolabeled gas from organic feedstocks, CAPyR suggests both oxidative and reductive chemistry from common Martian soil acquisitions. Metabolically active life combines these processes; a quantitative abiotic explanation of their combined heat labile response patterns remains to be demonstrated.

## 1. Introduction

Independent of any interpretation, the 1976 Viking Carbon Assimilation–Pyrolytic Release (CAPyR) experiment operationally detected the transfer of radioactive carbon, supplied as $^{14}CO_2$ and $^{14}CO$ gases, into a soil sample which, after excess radiolabeled gases were removed with a 120 °C flush, released radiolabel upon pyrolysis at 635 °C. A portion of the released radiolabel was trapped by an organic vapor trap (Chromosorb-P/CuO) operating at 120 °C. The trapped $^{14}C$-bearing material was then oxidized to $^{14}CO_2$ by the CuO, after which it emerged from the trap into a radiation detector. There, it was quantified in what came to be called "Peak 2" (Brown et al., 1978; Horowitz et al., 1972, 1977; Hubbard et al., 1970).

This workflow was extensively tested with terrestrial soils before flight. In more than 80 analyses of sterilized Terran soils, none met the investigators' criteria for a statistically significant response from the trapped Peak 2 fraction (Hubbard et al., 1970). Conversely, significant trapped Peak 2 radioactivity was found under illuminated or dark conditions only in Terran soils containing metabolically active organisms, including Antarctic dry-valley soils (Hubbard et al., 1970). Peak 2 was therefore to be interpreted as a signal of biological fixation of $^{14}CO_2$ into reduced organic material, subsequently fragmented by 635 °C pyrolysis into volatile products retained by the trap.

Under this interpretation, on Mars, the radioactivity in Peak 2 was to be assigned as arising from nonvolatile organic material synthesized biologically from the supplied $^{14}CO_2$ and $^{14}CO$, carbon autotrophically "fixed" during a 5-day incubation (Horowitz et al., 1977; Hubbard et al., 1970). The fixation was not limited to photosynthesis, although illumination was provided in most of the experiments. Any biological process that incorporated the supplied inorganic carbon-14 into material retained during the 120 °C flush, released at 635 °C, and trapped in the Chromosorb could contribute to the Peak 2 signal (Horowitz et al., 1972).

On Mars, many of the CAPyR experiments produced small, but statistically significant, net Peak 2 signals that were intended to "represent organic matter synthesized from $^{14}CO$ or $^{14}CO_2$ during the incubation" (Horowitz et al., 1977) (Table 1). Although this was originally interpreted as a sign of Martian biology (Jastrow, 1977), the CAPyR investigators themselves ultimately judged a biological origin unlikely, in part because the response appeared to be more thermally stable than expected for hypothetical Martian biology (Horowitz et al., 1977). The nonsignificant U2 and U3 measurements also require consideration of the documented VL2 counter leak and its possible effects on signal recovery (Section 2.3).

Klein's contemporaneous analysis of the Viking biology experiments came to favor nonbiological chemistry, but Klein and Horowitz both stated that the explanation for the apparent CAPyR organic synthesis remained obscure (Horowitz et al., 1977; Klein, 1978).

These analyses were reopened for interpretation by the Phoenix discovery of perchlorate salts in Martian soil, reported in 2009 (Hecht et al., 2009). Perchlorate has changed the interpretation of certain Viking measurements that were originally suggested to show the absence of life, especially those done by a gas chromatograph-mass spectrometer (GC-MS) system (McKay et al., 2025; Navarro-González et al., 2010), and possibly to indicate that the Martian soils were 'self-sterilizing' (Crawford et al., 2003; Tsapin et al., 2000). The new analyses discussing perchlorate have, however, focused primarily on three other Viking experiments: the Labeled Release (LR) experiment, the Gas Exchange (GEx) experiment, and the GC-MS experiment. This paper complements the set and reanalyzes data from the CAPyR experiment as its primary focus.

This reanalysis distinguishes three levels of inference: (i) what Viking directly measured, (ii) what molecular process the CAPyR design intended that measurement to represent, and (iii) what biological or abiotic mechanism could have generated the measurements.

Several chemical ambiguities motivate this reanalysis. Alkaline soils can retain labeled (bi)carbonates, and soil oxidants can transform fixed organics during pyrolysis. Perchlorate can also generate small, chlorinated products whose retention by the CAPyR trap remains insufficiently characterized. These processes, together with abiotic mineral chemistry, may alter the relationship between precursor chemistry and the measured Peaks 1 and 2.

## 2. Experimental Logic and Viking Results

### 2.1 CAPyR measurement principle and analytical workflow

The archetypal CAPyR experiment incubated approximately 0.25 $cm^3$ of Martian surface material for 120 hours in a chamber initially containing Martian atmosphere, to which a 92:8 (v/v) mixture of $^{14}CO_2$ and $^{14}CO$ was added (Brown et al., 1978; Horowitz et al., 1977) (**Fig. 1**). Standard early runs began at ~9.8 mbar of total pressure. Run temperatures across the mission spanned approximately 8 to 26 °C (Horowitz et al., 1977). Most runs were illuminated with a xenon lamp behind a Schott WG335 filter that suppressed wavelengths below approximately 320 nm. Dark runs without illumination were also performed (Brown et al., 1978; Horowitz et al., 1977).

After incubation, the soil was heated to 120 °C to vent the radiolabeled atmosphere (Horowitz et al., 1977). The soil was then heated at 625–635 °C to fragment, or pyrolyze, any organic material that had been fixed. Evolved gases were swept through a Chromosorb-P/CuO organic-vapor trap maintained near 120 °C. Radiolabel in $^{14}CO_2$, $^{14}CO$, and other volatile $^{14}C$-labeled products that were not trapped then moved as gases into a radiation counter, which registered a signal that came to be known as Peak 1 (Horowitz et al., 1977). It was recognized that Peak 1 radioactivity could have come from organic material fixed in the soil but then oxidized to $^{14}CO_2$ and/or $^{14}CO$ with oxygen atoms derived from the soil. Indeed, this was observed in preflight tests of the CAPyR architecture on Earth.

The material retained on the organic vapor trap was later released at 640 °C. Here, any trapped $^{14}C$-labeled organic fragments were oxidized by CuO embedded in the Chromosorb-P to give $^{14}CO_2$. This material passed into a radiation counter and was counted as Peak 2.

To manage any $^{14}CO_2/^{14}CO$ that might enter into the trapped pathway, Horowitz et al. empirically estimated the background term as Peak 2(0) = $28.8 + 2.84 \times 10^{-5} \times$ (Peak 1), with a standard deviation of 27 dpm (disintegrations per minute). Net Peak 2 radiolabel was calculated to be the measured Peak 2 minus Peak 2(0) (see **Fig. 1** in Horowitz et al., 1977). This empirical carryover subtraction is distinct from the correction for the VL2 counter leak discussed in Section 2.3.

Unlike the LR and (as planned) the GEx experiments, the CAPyR experiment added no organic nutrient mixtures. Nevertheless, incubation at 8–26 °C without a harsh cold cycle, increased pressure and $CO/CO_2$ ratio, and removal of light below 320 nm changed the samples' environment. Water-vapor addition in selected runs was a further perturbation.

To form organic products containing C–H bonds from $^{14}CO_2$, a source of hydrogen atoms and reducing equivalents is required. In Terran photosynthesis hydrogen typically comes from water, with notable exceptions discovered after Viking (e.g. Rey et al., 2006). Since the CAPyR designers recognized that water was likely to be scarce in the *Viking* soils, they arranged to add $H_2O$ vapor in some, but not all, of the CAPyR experiments (Horowitz et al., 1972). Approximately 80 µg was added per injection, resulting in 0.032 wt% added water in a 0.25 g sample, enough to reach saturation in the empty chamber at 20 °C. The actual humidity and soil water activity after the 80 µg injection would be considerably smaller, depending on partitioning among the gas phase, soil, and chamber surfaces. In C5, the combined water-addition, evaporation, and heating sequence also tested an abiotic explanation of Peak 2 (Horowitz et al., 1977).

In GEx, humidification from liquid nutrient solution held below the soil released $O_2$ without initially wetting the sample directly (Klein, 1978). While the CAPyR experiment did not measure $O_2$, both CAPyR and GEx humidification steps must be considered to be material perturbations of the natural environment. Indeed, the $O_2$ released in the GEx experiment, in some interpretations of the set of life detection experiments, is a candidate electron acceptor for the process that released radiolabeled carbon dioxide in the labeled release (LR) experiment that was run separately on Mars on a separate aliquot from a common soil acquisition. This possibility requires sufficient oxygen availability and reaction rates under LR conditions.

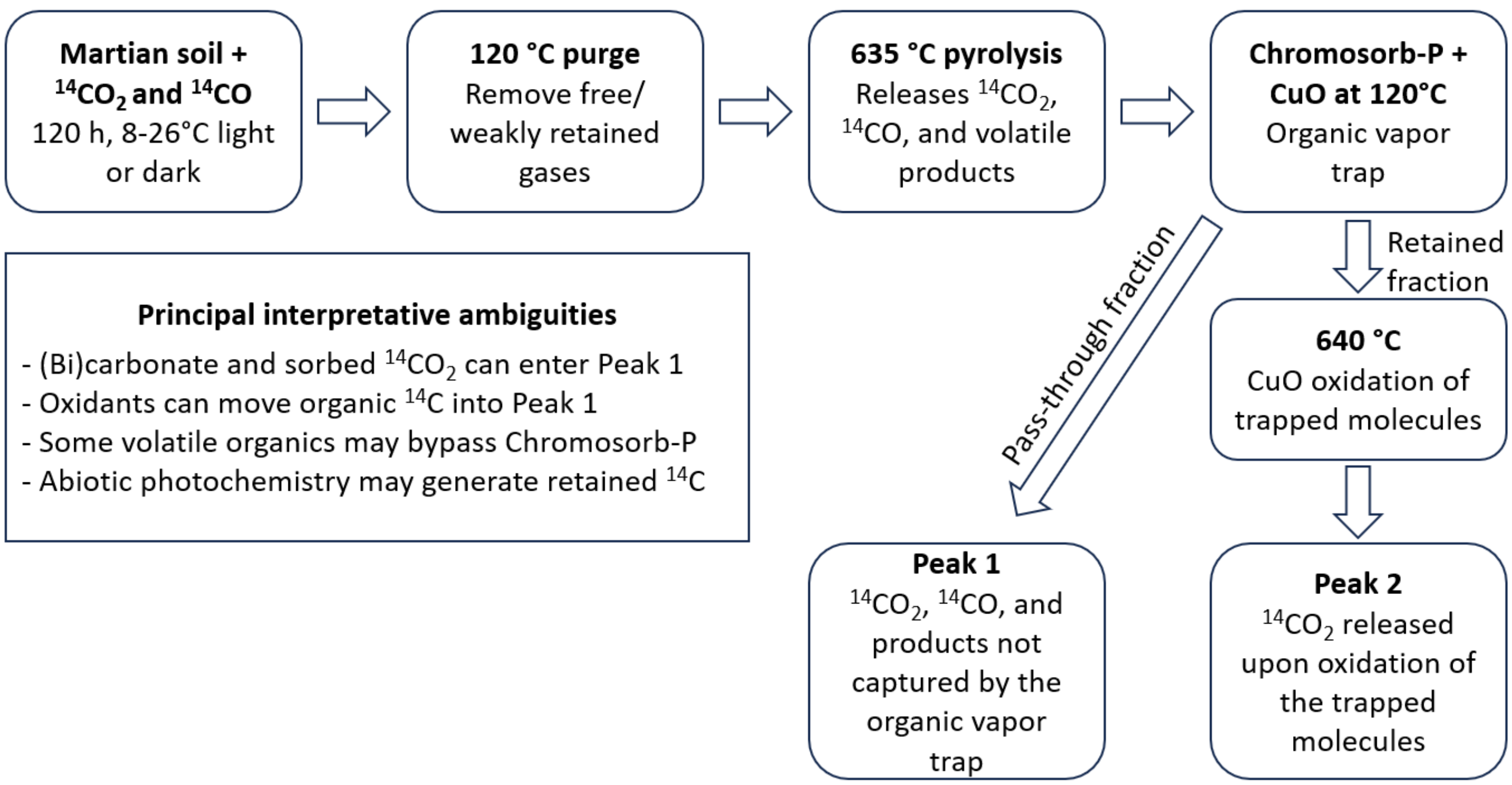

***Figure 1****. Viking CAPyR workflow redrawn for this reanalysis from CAPyR descriptions (Brown et al., 1978; Horowitz et al., 1977; Hubbard et al., 1970). Rounded boxes indicate analytical operations. Peak 1 and Peak 2 are operational fractions indicating the amount of $^{14}C$ counted in a beta particle detector, without molecular identification. The experimental design intended, however, for Peak 2 radioactivity to largely represent $^{14}C$-labeled organic fragments retained in the Chromosorb-P organic vapor trap, where those fragments were the pyrolysis products of organic molecules synthesized by reduction of the supplied $^{14}CO_2$ and/or $^{14}CO$.*

Preflight tests found picomole-scale analytical sensitivity to retained radiocarbon under the tested conditions in life-sparse Terran samples, routinely detecting metabolic activity of ~$10^2$ cells in the sample (Horowitz et al., 1972).

The design interpretation required that the pyrolysis-trapping-oxidation sequence deliver to net Peak 2 no radioactivity other than radioactivity from carbon that had traveled from $^{14}CO_2$/$^{14}CO$ into reduced organics and back to $^{14}CO_2$ by (first) pyrolysis into fragmented organics, (then) trapping the radioactive carbon as organic fragments on Chromosorb, (and then) subsequent oxidation to $^{14}CO_2$ by CuO (**Fig. 1**). Accordingly, careful preflight control experiments were carried out to constrain any contribution of unfixed $^{14}CO_2$/$^{14}CO$ to Peak 2 (Horowitz et al., 1972, 1977; Hubbard et al., 1971).

The design interpretation relied on distinguishing biological carbon assimilation from abiotic pathways capable of producing the same operational signal.

### 2.2 Preflight recognition of chemical ambiguities

Before flight, CAPyR developers investigated chemistry that could yield false-negative or false-positive interpretations (Hubbard et al., 1970). In nitrate-rich Atacama soil, oxidation during pyrolysis diverted radiolabel from reduced organic material into untrapped $^{14}CO_2$ and/or $^{14}CO$, increasing the Peak 1 pathway at the expense of the trapped fraction counted as Peak 2. These tests showed that soil oxidants could decrease recovery of fixed organic radiocarbon as Peak 2, as "the pyrolysis was converted to a partial combustion" (Hubbard et al., 1970). Nitrate oxidation was also considered when interpreting the failure of Viking GC-MS heating experiments to recover organic compounds (Biemann et al., 1976).

The Atacama test soil contained approximately 13 wt% nitrate, substantially more than the nitrate measured by Curiosity (Hubbard et al., 1970; Stern et al., 2015). In this respect, Atacama soil is not representative of Martian soil. However, these preflight data warned of the possibility that a soil oxidant might diminish the amount of Peak 2 radioactivity that might otherwise arise from pyrolysis of biological organics.

This warning became appreciated after Phoenix discovered perchlorate (as a salt) in Martian soils (Hecht et al., 2009). Perchlorate salts are comparatively unreactive toward organics at low ambient Martian temperatures. However, heating perchlorate-bearing mixtures, including at 635 °C, can oxidize organic material, primarily to $CO_2$ (that would partition to Peak 1), but also to traces of chlorinated organics (Navarro-González et al., 2010), including the $CH_3Cl$ and $CH_2Cl_2$ that were observed by the Viking GC-MS on Mars (Biemann et al., 1976, 1977).

Different preflight experiments showed that sterilized alkaline soil could retain large amounts of $^{14}CO_2$ and return that label upon pyrolysis to the $^{14}CO_2$ Peak 1 fraction while delivering very little (0.02% in the Atacama soil test) into the Peak 2 trapped-organic fraction (Hubbard et al., 1970). This established that Peak 1 could represent a substantial contribution from abiotic mineral-associated (bi)carbonates. Simple (bi)carbonate inorganic carryover into Peak 2 was small in the tested preflight matrices (Hubbard et al., 1970) and the calibrated background contribution to Peak 2 was estimated by the Peak 2(0) calculation (Horowitz et al., 1977) discussed above.

Peak 1 could therefore include radiolabel from both oxidation of pre-existing organic material and thermal decomposition of labeled (bi)carbonates that survived the 120 °C flush. Without further chemical discrimination, Peak 1 cannot be assigned uniquely to either source. An acidic-gas pretreatment, for example with HCl, could help distinguish acid-labile inorganic carbon by releasing it as $CO_2$ before pyrolysis. Such a treatment should be tested in a reconstructed CAPyR system to establish carbonate removal, preservation of relevant organics, and compatibility with the instrument and soil oxidants before being adopted for a future mission.

The team also considered Kolbe-type carboxylation, where $CO_2$ reacts without biological catalysis with electron-rich aromatic compounds to form radiolabeled benzoic acids, an abiotic route that might fix radiolabel from $^{14}CO_2$ into organic material. Experiments showed, however, that alkaline soil containing phenol or humic acid produced negligible amounts of trapped radiolabel that later generated Peak 2 radioactivity (18 and 3 counts per minute with 1% phenol and 10% humic acid, respectively). Hubbard et al. noted that this was because the pyrolysis would decarboxylate those benzoic acids in the reverse reaction, returning any $^{14}CO_2$ fixed by this path into Peak 1 $^{14}CO_2$ (Hubbard et al., 1970).

Lastly, preflight work had seen abiological synthesis of $^{14}C$-labeled material that might generate Peak 2 label from CO and water, driven by high-energy UV light (<300 nm) (Horowitz et al., 1977; Hubbard et al., 1971). This motivated the use of a lamp (as opposed to Martian sunlight) with a filter suppressing wavelengths below approximately 320 nm as the source of illumination when seeking photosynthetic carbon fixation (Horowitz et al., 1977; Hubbard et al., 1971). Hubbard later suggested that this spectral safeguard filter did not eliminate all abiotic sources of Peak 2 radiolabel (Hubbard, 1979), most notably demonstrated by the dark experiment. This suggestion is analyzed in Sections 3.2–3.5.

### 2.3 Run-level results and diagnostic treatments

Horowitz et al. (1977) reported ten CAPyR experiments, C1–C6 and U1–U4. Cycles with the prefix "C" were conducted by Viking 1 at Chryse Planitia (22.48° N, 47.97° W); those with the prefix "U" were conducted by Viking 2 at Utopia Planitia (47.97° N, 225.74° W). Nine cycles yielded results included in the quantitative summary; U4 was excluded because of an apparent valve failure (Table 1).

The VL2 measurements were complicated by a slow leak from the radiation counter, detected during analysis of U1 (Cooley, 1977: VI-15; Horowitz et al., 1976: 1322). This leak must be considered when evaluating the CAPyR data from this site.

Published records describe changes to the counting sequence and report that leakage corrections were applied. To guard against data loss should the leak worsen, the investigators changed the Peak 2 readout from sixteen-minute segments alone to both one-minute and sixteen-minute segments. Peak 1 was normally read out in one-minute segments (Horowitz et al., 1976). Cooley (1977) reported that the leak did not change significantly during the primary mission and that sequencing changes permitted improved leakage corrections. The report stated that corrections were calculated and applied and judged the science data valid, while acknowledging reduced confidence in the absolute radiocarbon levels.

U4 subsequently became unusable because an apparent valve failure allowed Peak 2 material to escape before counting (Horowitz et al., 1977). From helium consumption during the extended mission, Brown et al. (1978) inferred that valve S-11 had become stuck open, preventing collection of radiocarbon by pressurization of the detector volume. This extended-mission diagnosis does not quantify leakage during U1–U3. Cooper's later narrative described accelerated leakage when discussing U2 and progressively worsening leakage before U4 became unusable (Cooper, 1979b). The differing descriptions in Cooper and Cooley do not establish how the leak varied among U1–U3.

Although the primary-mission leak was acknowledged in the 1976 interim paper and Cooley's engineering report, the 1977 paper combining the analyses did not discuss the U1–U3 leak or its correction. The cited descriptions of the primary-mission leak do not provide cycle-specific leak rates, correction equations, or correction magnitudes. Nor do they explain quantitatively how uncertainty in the leakage corrections contributed to the reported errors. Consequently, these descriptions do not permit independent reconstruction of the corrections or assessment of their uncertainty.

Under the statistical analysis reported by Horowitz et al. (1977), C1–C6 and U1 yielded significant positive net Peak 2 responses, whereas U2 and U3 did not. Recovery of the Peak 2 fraction required transfer of radiolabel retained by the Chromosorb trap into the counter during heating of the trap to 640 °C. If not adequately corrected, sufficiently large radiocarbon losses before counting could reduce an otherwise detectable Peak 2 response below the significance threshold, producing a false-negative result.

Positive responses were observed in all three Chryse sample acquisitions represented by C1–C6. All three Utopia acquisitions represented by U1–U3 supplied results included in the quantitative summary, but only the first yielded a significant positive net Peak 2 response. Nonsignificant responses in U2 and U3 do not, by themselves, establish that the samples lacked Peak 2-generating activity.

The possibility that incompletely corrected radiocarbon losses contributed to the low Utopia responses must be considered in any assessment of the Utopia Planitia CAPyR experiments. This possibility is directly relevant to the magnitude of the U1 response and the interpretation of the nonsignificant U2 and U3 responses; the available documentation does not establish its quantitative importance.

Pending reanalysis of the original counting data and operating sequences, interpretations of these runs should acknowledge uncertainty associated with leakage corrections. This qualification is relevant to comparisons involving sample heterogeneity and illumination, and to reassessment of the earlier water-inhibition interpretation, which Horowitz et al. (1977) had already revised following C6. The authors are now in the process of obtaining primary data from repositories to resolve this underreported issue.

**Table 1. Run-level conditions and principal CAPyR results for the nine Mars experiments classified as usable in the original analysis. Values are from Horowitz et al. (1977). U4 is omitted because of an apparent valve failure.**

| Run | Sample / condition | Incubation T (°C) | Peak 1 (dpm) | Net Peak 2 (dpm) | Result |
|---|---|---|---|---|---|
| C1 | Sandy Flats 1; fresh; light, dry; no pretreatment | 17 ± 1 | 67,464 ± 536 | 842 ± 29 | Positive |
| C2 | Sandy Flats 1; stored 19 sols; light, dry; 175 °C × 3 h | 15 ± 1 | 69,536 ± 545 | 105 ± 29.5 | Positive |
| C3 | Sandy Flats 2; fresh; light, dry; no pretreatment | 13–26 | 61,027 ± 527 | 214.5 ± 28 | Positive |
| C4 | Sandy Flats 3; fresh; light, dry; no pretreatment | 16 ± 2 | 18,545 ± 381 | 289 ± 31 | Positive |
| C5 | Sandy Flats 3; stored 69 sols; light; $H_2O$ added/vented; 120 °C 1–2 min, then ~90 °C × 112 min | 17 ± 1 | 20,295 ± 395 | 275 ± 29 | Positive |
| C6 | Sandy Flats 3; stored 139 sols; light; $H_2O$; no thermal pretreatment | 15 ± 2.5 | 193,803 ± 864 | 255 ± 31 | Positive |
| U1* | Beta 1; fresh; dark, dry; no pretreatment | 15 ± 3 | 64,845 ± 527 | 178 ± 31 | Positive |
| U2* | Beta 2; fresh; light; $H_2O$; no pretreatment | 18 ± 1.5 | 113,845 ± 690 | −7 ± 28 | Not significant |
| U3* | Under Notch Rock; fresh; dark, dry; no pretreatment | 10 ± 2 | 118,309 ± 400 | 36 ± 35 | Not significant |

Note: “Positive” denotes statistical significance under the analysis reported by Horowitz et al. (1977); exact p values and the measured Peak 2 and Peak 2(0) components are given in the primary source. The U1–U3 results (*) were obtained with a documented counter leak; its residual effects on U2/U3 cannot be independently established from the published summaries (Section 2.3). Thus Peak 2 nonsignificance does not establish absence of the signal-generating agent.

The experiments were repeated under different conditions (Horowitz et al., 1977). The U1 Peak 2 radioactivity was substantial even though the incubation had been done without light. This suggests that a mechanism requiring illumination during the incubation cannot explain every positive Viking run (Horowitz et al., 1977).

Absence of light alone may not, however, exclude abiotic mineral chemistry in general. Hubbard (1979) reported some CAPyR-positive mineral responses from experiments done in the dark. As discussed below, the products, energy sources, and reaction pathways responsible for the dark responses remain unresolved.

C1 produced the largest net Peak 2, 842 ± 29 dpm (Horowitz et al., 1977). Horowitz et al. estimated that this corresponded to approximately 10 pmol of carbon fixed if derived entirely from CO, or approximately 30 pmol if derived entirely from $CO_2$ (Horowitz et al., 1977). The difference reflects the effective specific activities of $CO_2$ and CO in the incubation headspace, including dilution by the original Martian atmosphere. If interpreted biologically, that amount corresponds to approximately $10^3$ *Escherichia coli*-equivalents of carbon using an order-of-magnitude cellular carbon content. This is an illustrative mass-scale comparison (Phillips and Milo, 2009) and cannot be used to infer the initial population density without information about metabolic rates.

More information was sought by varying the CAPyR experimental conditions. Thus, the C2 sample, from the same dig as the C1 sample, was stored for 19 sols and then heated to 175 °C for 3 h. Its net Peak 2 was approximately 88% lower than fresh C1, while its Peak 1 was essentially unchanged (Horowitz et al., 1977). This contrast is consistent with a heat-sensitive contribution to Peak 2 and a different dominant source of Peak 1. Because storage and heating varied together, C1/C2 alone does not isolate their contributions. However, we assume the storage had little effect.

C4, C5, and C6 were aliquots from the same sample acquisition and gave similar net Peak 2 responses, when tested fresh, after 69 sols of storage plus the C5 water/thermal treatment, and after 139 sols of storage, respectively (Horowitz et al., 1977). These experiments differed in several variables, including radiolabeled-gas injections, storage, water addition, and heating histories.

The run histories constrain interpretations of thermal and storage sensitivity (Horowitz et al., 1977). In particular, the differences in Peak 1 radioactivities are noteworthy. The variability must be evaluated in light of both water history and the amount and specific activity of labeled gas supplied. C4, C5, and C6 received one, three, and six radioactive-gas injections, respectively (Horowitz et al., 1977; Brown et al., 1978). Injection counts do not establish a 1:3:6 molar dose ratio: reservoir pressure declined during the mission, and extra C5 injections partly compensated for that depletion. Horowitz et al. (1977) also associated C6's high Peak 1 with water retained during incubation, whereas C5's added water had been evaporated before labeling.

## 3. Reanalysis in Light of Post-Viking Chemical and Mineralogical Information

### 3.1 High-temperature oxidants can reduce Peak 2

The Mars Phoenix lander measured perchlorate ions after soil was added to an aqueous solution, corresponding to approximately 0.4–0.6 wt% $ClO_4^-$ in the soil and indicating the presence of perchlorate salts (Hecht et al., 2009). While perchlorate salts are comparatively unreactive toward organics at low temperatures, heating these mixtures, often above approximately 400 °C, can oxidize organic material to generate $CO_2$ and small amounts of chlorinated products (Navarro-González et al., 2010), including the methyl chloride ($CH_3Cl$) that was observed by the Viking GC-MS (Biemann et al., 1976; Glavin et al., 2013; Guzman et al., 2018).

Perchlorate was not directly measured at either Viking landing site, so its abundance in the CAPyR soil samples remains unknown. Controlled experiments show that perchlorate-driven combustion of organic matter to $CO_2$/CO and substrate-dependent degradation/chlorination deliver a range of chlorinated organic compounds (Millan et al., 2020; Royle et al., 2018). However, the ability of Chromosorb-P to trap small volatile chlorinated organics (e.g. $CH_3Cl$) at 120 °C (the temperature at which the trap was operated on Mars) remains unknown. Different organics with different structures can generate different ratios of $CO_2$, small chlorinated organics (which may not be trapped by Chromosorb at 120 °C), and less volatile chlorinated organics (whose retention by Chromosorb at 120 °C also requires measurement) (Sephton et al., 2014). Studying these differences is a key part of any future work on Earth to add value to the Viking data without requiring another expensive mission to Mars.

These experiments establish high-temperature oxidation and chlorination of organic carbon that may have caused an underestimation of the amount of biologically fixed carbon. Under the CAPyR workflow, perchlorate chemistry that has repartitioned some pre-pyrolysis reduced $^{14}C$ towards the $^{14}CO_2$/$^{14}CO$ and volatile chlorinated organics entering Peak 1 and away from Peak 2 should be expected. The magnitude of that repartitioning remains unknown until measurements are made in a reconstructed CAPyR system, including a detailed analysis of how different organic molecules give different ratios of $CO_2$, very volatile organic products that are not trapped on Chromosorb, and less volatile organic fragments that are trapped.

The preflight experiments with soils containing nitrate demonstrate this principle and establish the direction of the effect (Hubbard et al., 1970). With nitrate as well as perchlorate, net Peak 2 could underestimate reduced autotrophically fixed labeled carbon present immediately before pyrolysis. Again, this was considered for nitrate by those analyzing the GC-MS data (Biemann et al., 1976).

High-temperature oxidation of organic material does not, by itself, explain formation of a labeled reduced precursor during the preceding incubation. This distinction does not exclude abiotic contributions to Peak 2: Hubbard (1979) investigated mineral-associated responses during incubation, whereas Quinn and Pacheco (2013) proposed that retained labeled inorganic carbonate could yield organic products during thermal analysis. These mechanisms concern different stages of the CAPyR workflow and are assessed separately below.

The essentially unchanged Peak 1 in C1 versus C2 also provides an important constraint. If much of the C1-to-C2 decrease in Peak 2 was caused by the 175 °C treatment, most of Peak 1 cannot plausibly have been generated by the same heat-labile agent. Conversely, Peak 2 cannot be assumed to contain all reduced carbon that may have existed before pyrolysis because perchlorates and nitrates can divert pre-existing reduced carbon toward the Peak 1 pathway during heating (Navarro-González et al., 2010; Royle et al., 2018). Given that Peak 1 was much larger than net Peak 2 in every run (**Table 1**), even substantial repartitioning of radiocarbon that would otherwise have contributed to Peak 2 into the Peak 1 pathway would produce only a small fractional increase in Peak 1.

Post-Viking GC-MS reanalysis provides further context. Guzman et al. (2018) in addition to chloromethane and dichloromethane identified chlorobenzene in the Viking data and interpreted it as consistent with a Martian aromatic-carbon source transformed during analysis. Such high-temperature products can contain indigenous carbon without uniquely identifying the original precursor structure. Laboratory release of chloromethane from carbonaceous-meteorite material likewise illustrates precursor ambiguity (Keppler et al., 2014). These findings weaken the inference that failure to recover parent organic molecules implies their absence before heating (Benner et al., 2000; Guzman et al., 2018; Navarro-González et al., 2010).

### 3.2 Laboratory abiotic chemistry reproduced only a subset of CAPyR results

Immediately after Viking, Hubbard (1979) used Martian soil simulants to explore the possibility that the outcomes of the CAPyR experiments might have arisen without biology. These experiments used a Test Standards Module (TSM) with mechanical components and operational sequences similar to the flight CAPyR system. This included a WG335 filter that removed wavelengths from a standard lamp below approximately 320 nm (Hubbard, 1979).

With TSM, Hubbard (1979) obtained Peak 2 radiolabel with hematite, maghemite, magnetite, a clay-rich Mars-analog mixture, and Fe-exchanged montmorillonites. This suggested that selected mineral substrates might generate CAPyR-positive analytical responses without biology.

However, the responsible chemistry was not uniform or fully resolved. For the clay-rich Mars analog and Fe-exchanged montmorillonites, Hubbard favored photocatalytic synthesis or isotopic exchange involving $^{14}CO$ and surface-bound carbon or hydroxyl groups. Magnetite samples incubated in the dark under dry conditions produced slightly larger Peak 2 responses than corresponding illuminated samples, leading Hubbard to state that a photochemical process was contraindicated. Hubbard did not identify the reaction responsible for this "dark magnetite" signal. Thus, Hubbard showed an abiotic dark Peak 2 response but did not establish how $^{14}CO_2$ and/or $^{14}CO$ acquired the chemical form responsible for that signal. In particular, Hubbard did not establish the energy and redox balances or identify the products responsible for this dark response (Hubbard, 1979).

Evaluating these mechanisms requires the chemical identity of the labeled products and a model of energy and redox balance. The wavelength cutoff at 320 nm significantly constrains the photochemical pathway. Further, a dark response requires a nonphotochemical explanation, which could involve chemical disequilibrium, isotope exchange, or transformations during subsequent analysis. Thus, the abiotic observations reported by Hubbard (1979) will remain paradoxical until the responsible chemistry is shown to satisfy redox and thermodynamic constraints, possibly requiring more Earth-based studies.

Hydration of CO to form formic acid, $CO + H_2O \rightarrow HCOOH$, does not change the carbon oxidation state. However, uncatalyzed gas-phase hydration is unlikely to produce a measurable Peak 2 response. Establishing a surface-mediated pathway requires demonstration of product formation, retention of labeled carbon through the 120 °C flush, and recovery in Peak 2 after pyrolysis and Chromosorb-P trapping.

A mechanism producing carbon more reduced than that in CO must identify the compensating oxidation reaction or other source or sources of reducing equivalents. The possible involvement of another CO molecule, reduced mineral constituents, or pre-existing organics cannot be excluded solely because no $H_2$ was added. Hubbard's results establish abiotic analytical responses while leaving their complete mechanism unresolved.

Hubbard's experiments demonstrate (Claim A) that selected mineral systems can generate abiotic positive Peak 2 responses under some conditions (Hubbard, 1979). However, they did not establish Claim B: that the abiotic chemistry explains the complete set of CAPyR signals in the Viking regolith (Hubbard, 1979). Accordingly, Hubbard concluded that duplication of the whole of the Viking CAPyR data had not been achieved (Hubbard, 1979).

Specifically, several candidate soil substrates produced Peak 1 values much larger than Viking, perhaps involving generation of $^{14}C$-labeled carbonates and bicarbonates. Further, responses to water and thermal treatment differed among substrate systems (Hubbard, 1979). Hubbard (1979) recognized another issue: Moisture increased the TSM response in some Hubbard experiments, whereas the Viking experiments showed no reproducible moisture effect. Horowitz et al. (1977) ultimately favored sample heterogeneity over inhibition by the increased humidity as the explanation for the differing Utopia responses to vapor addition. The documented VL2 counter leak further limits the use of U2/U3 to distinguish water inhibition from sample heterogeneity (Section 2.3; Horowitz et al., 1976; Cooley, 1977).

Thus, it remains open whether an abiotic process or abiotic processes can reproduce the Viking CAPyR data as a whole. This open question is separate from the larger question as to whether a combination of non-biological oxidation and non-biological reduction processes, in one soil, can account for the oxidation seen in the labeled release (LR) experiment and the reduction seen in the CAPyR experiment (see below).

### 3.3 Mineralogical applicability and the partially circular analog argument

Earth-based laboratory work to analyze the Viking data has begun with a choice of material to simulate Martian soil. Hubbard identified model-soil selection as the aspect of the simulations in which he had the least confidence. The then available Viking X-ray fluorescence data constrained elemental abundances, but did not assign mineral phases (Hubbard, 1979).

Recognizing this ambiguity, Hubbard used a 'Mars analog soil' that contained 51% nontronite and 26% bentonite, making it 77% clay of some type (Hubbard, 1979). Later $Fe^{2+}$- and $Fe^{3+}$-exchanged montmorillonites were considered particularly promising, because they produced positive Peak 2 with Peak 1 values closer to the Viking range (Hubbard, 1979).

The similarity of the analytical output being explained, the $^{14}C$-label in Peak 1 in particular, was used in part to support the geological realism of the substrate chosen to explain that output (Hubbard, 1979). Analytical agreement supports a chemical model, but its geological applicability requires independent evidence for the substrate. Otherwise, one substrate is selected from many simply because it generates the data to be fitted.

Modern mineralogical measurements provide independent constraints that were unavailable in 1979. Thermal infrared analysis of Martian atmospheric dust indicates a framework of silicate, probably feldspar, as a major component, with lesser olivine, pyroxene, amorphous material, hematite, and magnetite. This differs from a composition that is dominated by clays (Hamilton et al., 2005).

Rocknest data provided the first *in situ* X-ray diffraction measurement on Mars to identify mineral phases. Direct X-ray diffraction of the Rocknest aeolian soil identified plagioclase, olivine, augite, pigeonite, minor iron-bearing phases, and a substantial amorphous component, with a crystalline assemblage similar to basaltic materials (Bish et al., 2013).

No corresponding measurements constrain the minerals in the Viking Chryse or Utopia samples. Thus, Rocknest observations do not eliminate the possibility that the Viking soils contain smectites. However, they do not provide independent support for treating a 77%-clay or artificially Fe-exchanged montmorillonite substrate as a default representation of Viking surface fines (Bish et al., 2013; Hamilton et al., 2005). The mineralogy weakens the claim that the abiotic chemistry observed with these soil simulants explains the complete set of CAPyR signals in the Viking regolith (Section 3.2).

### 3.4 Thermal response and storage behavior favor a heat-labile agent over a Hubbard 1979 model

While the products and energy and electron balances of these mineral reactions remain unresolved, we can compare biological and abiological models for their ability to explain the variants of the Viking CAPyR experiments. The C1/C2 comparison provides the strongest thermal attenuation in the Viking data that might be interpreted as the consequence of biology. C1 gave 842 ± 29 dpm net Peak 2, whereas C2 gave 105 ± 29.5 dpm net Peak 2 after 19 sols of storage plus 175 °C heating for 3 h, an ~88% decrease (Horowitz et al., 1977).

Because storage and heat treatment were varied together, the entire decrease cannot be assigned to the 175 °C heating. For comparison, C4, C5, and C6, aliquots from one Sandy Flats 3 acquisition, gave similar net Peak 2 after 0, 69, and 139 sols of storage, despite additional treatment differences and differences in the number of radiolabeled-gas injections, suggesting a little contribution of the storage to the C2 response (discussed in **Section 2.3**).

The comparison also constrains attempts to explain CAPyR and the Viking Labeled Release (LR) response (the oxidation of $^{14}C$-labeled organics to give $^{14}CO_2$). Klein's cross-experiment summary reported that the Peak 2 CAPyR signal was retained after prolonged storage, while the LR response disappeared. The samples in that comparison were stored at spacecraft temperatures of approximately 20 °C (Klein, 1978).

Brown et al. (1978) noted that the biology instrument used a common Soil Distribution Assembly that retained sufficient material from each acquisition for subsequent redistribution to the GEx, CAPyR, and LR experiments. However, publications do not establish identical microscopic humidity or headspace histories for every stored aliquot. The comparison is therefore best described as a broadly common dry spacecraft-temperature storage environment.

Under that shared-regime assumption, the different LR and CAPyR responses argue against the same labile, storage-sensitive agent being responsible for both the oxidation (LR) and reduction (CAPyR) seen in the same partitioned sample (Jastrow, 1977). The observations summarized by Klein (1978) are consistent with agents or mechanisms with different thermal and storage stabilities (**Fig. 2**). The CAPyR Peak 2-generating activity remained detectable after 139 sols of storage, was not detectably lower in C5 than in C4 following a brief excursion to 120 °C followed by approximately 90 °C for 112 min, and remained significantly above background after the C2 treatment at 175 °C for 3 h. In contrast, the rapid LR nutrient-oxidizing response was absent after ~80 sols of storage, was partially attenuated by 50 °C for 3 h, and was eliminated by 160 °C for 3 h. Here, however, the LR experiments added bulk water, while the CAPyR experiments did not.

Hubbard's mineral systems also did not reproduce the Viking CAPyR thermal attenuation (Hubbard, 1979). Several candidate substrates were pre-dried at 160 °C before testing, while others were dried over $P_2O_5$. Nevertheless, substantial abiotic activity remained (Hubbard, 1979). In the Test Standards Module, a simulated sterilization treatment of 170 °C for 3 h retained 72.5% of the corresponding light-dry gross Peak 2 response of the Mars analog, and sterilized maghemite did not show a Viking-like reduction in Peak 2 (Hubbard, 1979).

Hubbard therefore cautioned that the sterilization effect was not statistically resolved in his limited analog series, and that the tested mineral systems did not exhibit the strong attenuation observed between Viking C1 and C2 (Hubbard, 1979). For the CAPyR data, Klein had already noted that a successful nonbiological catalyst would need to remain active after heating at approximately 90 °C yet be strongly impaired by 175 °C (Klein, 1978).

The Viking thermal response therefore favors an unknown heat-labile source of Peak 2 over the mineral-catalyzed abiotic mechanisms explored by Hubbard (1979). This is consistent with biology, especially recognizing that some Terran life forms can be recovered in viable forms after heat treatment comparable to Viking C2 (Kempf et al., 2008; Schubert and Beaudet, 2011). It shows that at least some known life forms may survive the 175 °C for 3 h treatment and conceivably cause a similar response to the one observed in C2. This finding does not exclude any (yet to be discovered) abiotic carbon-fixing pathways with demonstrated thermal susceptibility. Finally, a mixed explanation could assign the residual C2 signal to abiotic chemistry while allowing biological contributions to the unheated samples.

In C5, two injections of water vapor were added to the sample. After 4 h of low temperature incubation, the chamber was vented and the sample was heated to 120 °C for 1–2 min, then held near 90 °C for approximately 112 min (Horowitz et al., 1977). The temperature was subsequently lowered to 17 ± 1 °C, after which three injections of radiolabeled $CO_2$/CO were added for the subsequent incubation (Horowitz et al., 1977). Net Peak 2 did not differ significantly from that of fresh C4 (Horowitz et al., 1977).

The C5 essentially dry heat pretreatment cannot be considered sterilizing by terrestrial standards. Vacuum-dried wild-type *Bacillus subtilis* spores retained 55% viability after 5 min at 120 °C, while 90% remained viable after 60 min at 90 °C (Setlow and Setlow, 1995). Thus, it is conceivable that at least some Terran life forms would yield a similar response to that observed in C5: robust survival or persistence of the carbon-assimilation machinery during heating followed by activity during the cooler incubation.

Anhydrobiotes, organisms that tolerate essentially complete dehydration, are known to withstand very high temperatures sometimes well above temperatures found in nature (Ciera et al., 2014; Mertens et al., 2008). Such an exaptation, or the emergence of traits not directly selected for, may be expected on Mars. Thus, given the current dry conditions on Mars, the persistence of Peak 2 observed in C5 and diminished yet retained signal in C2 does not exclude a biological origin of the signal. This does not imply that Martian life would resemble known bacterial spores.

### 3.5 Other proposed abiotic routes remain unvalidated

***3.5.1 Kolbe-type carboxylation***. Preflight tests examined Kolbe-type carboxylation, in which $^{14}CO_2$ could be incorporated into phenolic organic matter without biological catalysis (Hubbard et al., 1970). Alkaline soil supplemented with phenol or humic acid yielded little radiolabel in the trapped-organic fraction after pyrolysis (Hubbard et al., 1970). The investigators attributed this low retention to the expected release of carboxyl carbon as $CO_2$ during pyrolysis, so this route did not produce a trapped-organic false positive in the tested matrices (Hubbard et al., 1970).

***3.5.2 High-temperature reduction*.** The original Viking GC-MS experiment was designed to explicitly avoid having $H_2$ contact the Martian soil. Instead, it used $^{13}CO_2$ to flush the heated organics from the Martian soil. Biemann et al. (1977) wrote that their "use of $CO_2$ rather than $H_2$, which is the carrier gas for the gas chromatograph, avoids catalytic or thermally induced reduction of the organic material possibly present in the sample" (Biemann et al., 1977). However, VL-2 used hydrogen expansion during the principal heating series for both samples, followed by a final reheating in $^{13}CO_2$ (Biemann et al., 1977). As the high temperature radiocarbon reduction with $H_2$ was known to the team, CAPyR used a helium stream, with no supplied $H_2$ (Horowitz et al., 1977).

In a 2013 conference abstract, Quinn and Pacheco proposed that labeled inorganic carbonates could form during low-temperature incubation and subsequently yield organic products during thermal analysis (Quinn and Pacheco, 2013). Thus, their proposal does not require a reduced organic precursor to form before the pyrolysis step. They reported trace $CH_3Cl$ and $CH_2Cl_2$ during heating of metal carbonates and hydrated perchlorate mixed with magnetite and hematite and cited Jagadeesan et al. (2009) as evidence for carbonate reduction under hydrogen at high temperatures.

Jagadeesan et al. (2009) demonstrated methane formation during hydrogen treatment of mixed metal carbonates, with reduced transition metals and oxide nanoparticles acting as catalysts. In a later study Jagadeesan et al. provide more details, showing the hydrogen reduction of iron-oxide/calcium-carbonate mixtures at ~400 °C produced up to 23% $C_1$–$C_3$ hydrocarbons (Jagadeesan et al., 2013). These studies demonstrate carbonate reduction under an externally supplied 1-5 atm hydrogen atmosphere. However, those conditions are far from the conditions during CAPyR, where heating was done under a dry He atmosphere.

However, in the same abstract Quinn and Pacheco (2013) also noted detection of "trace levels of $CH_3Cl$ and $CH_2Cl_2$ during the thermal volatilization of metal carbonates and hydrated perchlorate mixed with magnetite and hematite" when using "He in head space and carrier gas." The helium experiments leave the sources of hydrogen atoms and reducing equivalents unresolved. Water released from hydrated salts may supply hydrogen, but carbonate reduction would require an identified electron donor and a favorable overall reaction. Neither the reaction mechanism nor its thermodynamic feasibility under the reported conditions was established. No follow up "experiments designed to confirm this observation and to determine relevant mechanisms" (Quinn and Pacheco, 2013) were found by the authors. Retention of water by the hydrated perchlorates during the 120 °C venting step done prior to Viking pyrolysis should be also analyzed.

Since the abstract is not clear on the labeling of the observed traces of $CH_3Cl$ and $CH_2Cl_2$, chlorination of pre-existing organic contamination may provide an alternative explanation for the reported observation. The chlorination was demonstrated with acetone in the same study and with soil organics by Navarro-González et al. (2010).

Although authors of this reanalysis are skeptical of the chemical plausibility of high temperature reduction of carbonates with hydrated perchlorates suggested by Quinn and Pacheco (2013), Mars may have provided another source of hydrogen to support at least trace amounts of reduced organics following the high temperature $H_2$ reduction of carbonates.

Some $H_2$ could have been generated from Martian soil during CAPyR's post-flush pyrolysis. Curiosity's SAM instrument detected $H_2$ from heated Martian materials, with proposed sources including reactions of $Fe^{2+}$-bearing phases with evolved water and decomposition of refractory carbon (Sutter et al., 2017). Rocknest's reported total corresponds to approximately 60 µmol $H_2$ per gram of soil, with a release feature near 600 °C. However, the total covers heating from 40 to approximately 870 °C at 35 °C $min^{-1}$ and the authors attribute the >800 °C $H_2$ release to an instrument effect. Water fragmentation in the mass spectrometer also contributes to some lower-temperature signals. The 120 °C pre-pyrolysis flush in Viking CAPyR would reduce the amount of available water for the SAM's proposed branch of the $H_2$ production.

Even assuming similar $H_2$-producing properties at the Viking sites, these measurements do not establish the yield during CAPyR's approximately 635 °C pyrolysis, whose specifications called for a heat-up of less than 90 s followed by a 90 s hold (Horowitz et al., 1977; Brown et al., 1978).

The high temperature carbonate reduction with locally produced $H_2$ is a plausible pathway towards an abiotic explanation of the Peak 2. The demonstration should include the preceding 120 °C flush and a test of heat sensitivity, and demonstrate reaction with retained $^{14}C$ and recovery of the resulting products in the trap to produce Peak 2. Note that methane would not be captured by Chromosorb-P, and retention of $CH_3Cl$ on Chromosorb-P is yet to be established.

***3.5.3 Low-temperature $^{14}C$ reduction with atmospheric $H_2$.*** Free atmospheric $H_2$ would be removed during venting and the helium purge before pyrolysis; however, the atmospheric $H_2$ was present during the incubation. We therefore suggest a new pathway based on this observation. An illustrative atmospheric inventory can be calculated from $n(H_2) = x(H_2)PV/RT$. A 4 $cm^3$ gas volume (Brown et al., 1978) at 7.7 mbar and 0 °C, with 15 ppm $H_2$ (Krasnopolsky and Feldman, 2001), contains approximately 20 pmol $H_2$. For reduced carbon with formaldehyde-like elemental composition, reduction of one mole of CO requires one mole of $H_2$ per carbon whereas reduction of $CO_2$ requires two: the C1 10 pmol CO or 30 pmol $CO_2$ end-member estimates would therefore require approximately 10 and 60 pmol $H_2$, respectively, before accounting for other metabolic or chemical demands. Thus, a yet unknown abiotic reduction process could have plausibly used the atmospheric $H_2$ to reduce CO to yield the CAPyR C1 response. This hypothesis also provides roughly stoichiometric constraints on the magnitude of the observed Peak 2 responses. Due to slow kinetics of abiotic radiocarbon fixation at low-temperature and low-pressure, a yet unknown efficient mineral catalyst would be required that operates in the dark and is strongly attenuated at 175 °C.

On the other hand, low-temperature trace gas metabolism is known in biology. For example, Ji et al. (2017) showed aerobic oxidation of atmospheric $H_2$ and CO coupled to chemosynthetic $CO_2$ fixation in Antarctic soil communities. This establishes a biological precedent for trace-gas-supported carbon fixation occurring at Mars-relevant temperature. We may hypothesize that Martian organisms could have evolved to use $H_2$ to reduce CO into hydrocarbons or carbohydrates, as this would be more energetically favorable than known $CO_2$ fixation with $H_2$ (Rey et al., 2006).

### 3.6 Summary: Biology is consistent with CAPyR observations, while a complete abiotic explanation remains unresolved

Abiotic carbon chemistry remains relevant. Candidate routes include incubation-stage mineral chemistry and reduction of retained labeled carbonates during pyrolysis, potentially supported by sample-generated $H_2$ (Hubbard, 1979; Quinn and Pacheco, 2013; Sutter et al., 2017). Fe(III) is not itself an electron donor in its usual reduction chemistry, whereas Fe(II) in magnetite could supply reducing equivalents. Whether such chemistry accounts quantitatively for the CAPyR responses, including isotope incorporation and trap recovery, remains unresolved.

No demonstrated abiotic mechanism or combination of mechanisms has yet reproduced the combination of Viking Peak 1/Peak 2 behaviors, the light/dark results, the C5 response to heat treatment, and the strong C2 attenuation after heat treatment. Because sample acquisitions and experimental conditions differed, a successful model need not reproduce every run identically. It should generate positive net Peak 2 signals and reproduce the principal classes of observations and diagnostic treatment responses. The nonsignificant U2/U3 results should be evaluated with the instrumental uncertainty described in Section 2.3, without assuming that they require an absence or attenuation of the responsible chemical or biological agent. Likewise, due to the ongoing leak in the system, the lower Peak 2 in U1 cannot be solely ascribed to the absence of illumination.

A biological contribution remains a plausible interpretation of the CAPyR data. Under the original interpretation, a Martian biological system would need to incorporate supplied $^{14}CO_2$ and/or $^{14}CO$ into a retained reduced-carbon precursor during the 5-day incubation at 8 to 26 °C (Horowitz et al., 1977). Its relevant carbon-assimilation activity would need to survive the C5 120 °C/90 °C pretreatment while accounting for the attenuation after C2's combined storage and 175 °C treatment (Horowitz et al., 1977). A mixed explanation could assign the residual C2 signal to abiotic chemistry.

Survival following exposure to the C5 temperature range is compatible with some terrestrial spores (Kempf et al., 2008; Schubert and Beaudet, 2011; Setlow and Setlow, 1995). Likewise, the significant U1 result, obtained in the dark, is compatible with nonphotosynthetic carbon fixation, including atmospheric trace-gas-supported chemolithoautotrophy (Ji et al., 2017). CO/$H_2$-based metabolism could support both carbon assimilation and energy conservation. Depending on the pathway and conditions, CO/$H_2$ metabolism might also yield energy and/or water.

The CAPyR data neither identify a metabolic pathway nor prove the presence of living cells. They show only that a Martian biological catalyst with this activity profile constrained by the reported treatments is consistent with the Viking observations. An abiotic mechanism or combination of mechanisms reproducing the combined illumination, thermal, and storage response pattern remains undemonstrated.

## 4. Future Earth-Based Testing and Future Missions

This reanalysis motivates reconstruction of the CAPyR experiments on Earth, without requiring a new flight to Mars. Experiments with hematite, maghemite, magnetite, original clay-rich Mars analog, and Fe-exchanged montmorillonites should be replicated as historical positive controls because they establish a range of possible abiotic CAPyR-positive responses (Hubbard, 1979). Their products, reaction pathways, and energy and electron balances should then be established. In particular, the hydration of carbon monoxide to form formic acid, whose thermodynamics depend on the reaction conditions, should be re-investigated as a mechanism to create small amounts of a Peak 2 signal.

In parallel, substrates should be selected using modern models of Mars mineralogy, including basalt-derived plagioclase/pyroxene/olivine assemblages, amorphous Fe-bearing material, measured minor iron oxides, and systematically varied clay abundances (Hamilton et al., 2005; Bish et al., 2013).

The reconstruction should then reproduce the Viking CAPyR workflow, including gas composition, illumination spectrum, incubation, the 120 °C flush, high-temperature sample pyrolysis, Chromosorb-P/CuO trap, and oxidation by CuO in the trap (Horowitz et al., 1977; Brown et al., 1978). Each experiment should be subjected to light/dark controls, storage-only controls, a C5-like thermal treatment, and a 170–175 °C three-hour treatment, with Mars-relevant perchlorate added as a separate variable. (Bi)carbonate and Kolbe-type routes can be examined in the same system.

Further, any proposed chloromethane pathway should be tested with isotope tracing to establish carbon provenance and direct measurement of retention of the proposed and other highly volatile chlorinated organics by the CAPyR trap. A candidate mechanism should be judged against both Peak 1/Peak 2 generation and attenuation as a treatment response, noting what fraction of organics oxidized by perchlorate is moved from Peak 2. Producing Peak 2 alone is insufficient to reproduce the Viking pattern.

The Utopia uncertainty also calls for recovery and reanalysis of time-resolved counting records, command sequences, engineering measurements, and the original leakage corrections. Comparisons across U1–U4 should distinguish gas loss during collection from gas loss during counting and test whether the available corrections account for the low U2/U3 responses. In a reconstructed CAPyR instrument, controlled-leak experiments could quantify the effects of incomplete valve sealing on both peaks under the original and revised counting sequences (Brown et al., 1978; Cooley, 1977; Horowitz et al., 1976).

Future missions using the CAPyR concept could test an acid-vapor pretreatment to distinguish acid-labile inorganic carbon, provided its selectivity and compatibility are established. Direct measurements of retained radiocarbon without pyrolysis, combined with validated removal of unassimilated inorganic label, would provide a complementary approach. Separate CO and $CO_2$ incubations could identify the carbon source more clearly. With these updates, CAPyR-style experiments are good candidates for future Mars astrobiology search. Such designs merit evaluation for compact platforms delivered in swarms, including IMPRESS penetrators (Spacek et al., 2026).

## 5. Conclusions and Overview

The CAPyR reanalysis can be summarized in six technical points.

1. CAPyR was operationally designed and preflight-tested to detect the transfer of $^{14}C$ from supplied $^{14}CO_2$ and $^{14}CO$ into a radiolabeled fraction (Peak 2) that, under the intended molecular interpretation, represented biologically formed organic material synthesized by a reduction reaction in an analog of Terran autotrophy (Brown et al., 1978; Horowitz et al., 1972; Hubbard et al., 1970).

2. Viking measured statistically significant positive net Peak 2 signals in seven of nine runs classified as usable under the published analysis: all six Chryse runs and the dark U1 run (Horowitz et al., 1977). Taken at face value, this pattern suggests significant responses in all three Chryse acquisitions but only in the first Utopia acquisition. However, a counter leak was documented from U1 onward, with counting-sequence changes and corrections applied during the primary mission (Cooley, 1977; Horowitz et al., 1976). The available published summaries do not establish whether residual leakage contributed to the nonsignificant U2/U3 signals, and the later S-11 failure does not prove an earlier fault in that valve (Brown et al., 1978). Thus, seven of nine remains a description of the published results; U2/U3 should be interpreted cautiously as evidence about the presence or distribution of the responsible chemical or biological agent. The positive responses met the reported statistical criterion for net Peak 2, and the C1/C2 comparison shows substantial attenuation after the combined storage and heat treatment (Horowitz et al., 1977).
3. Preflight experiments with nitrate-containing soils directly showed that an oxidizing agent in a soil sample can, during pyrolysis, move reduced $^{14}C$ away from the Peak 2 fraction captured by the organic vapor trap into $^{14}CO_2$ in Peak 1 (Hubbard et al., 1970), undercounting fixed carbon. Later experiments independently demonstrate oxidation and chlorination of organic material during heating with perchlorate (Millan et al., 2020; Navarro-González et al., 2010; Royle et al., 2018). Under the Viking CAPyR workflow, analogous chemistry could therefore cause Peak 2 to underestimate the amount of radiolabeled organics formed by reduction of $^{14}CO_2$ or $^{14}CO$. This repartitioning has not yet been quantified within a complete CAPyR workflow.
4. *Oxidation* of pre-existing organic matter during pyrolysis does not, by itself, explain formation of a labeled *reduced* precursor during incubation. However, abiotic hypotheses include both incubation-stage mineral chemistry and formation of labeled organic products during thermal analysis of retained inorganic carbon (Hubbard, 1979; Quinn and Pacheco, 2013). The available studies do not establish a quantitative reconstruction of the CAPyR observations as a whole.
5. Modern Mars chemistry and mineralogy weaken two premises that once favored a nonbiological interpretation of the Viking results. In the presence of perchlorate, high-temperature analysis can destroy or transform pre-existing organics. Thus, any failure, by the GC-MS or CAPyR, to recover parent organics or their fragments does not establish that none were present before heating (Guzman et al., 2018; Navarro-González et al., 2010). At the same time, the clay systems used to seek non-biological results for the CAPyR data (Hubbard, 1979) were selected under a mineralogical model not independently supported as a default composition of Martian aeolian fines. Further, even these systems did not reproduce the thermal attenuation and other details of the CAPyR observables.
6. If the C1 net Peak 2 carbon arose from biological assimilation, its magnitude corresponds to approximately 10 pmol if derived from CO or 30 pmol if derived from $CO_2$, on the order of $10^3$ *E. coli*-equivalents of carbon in mass (Horowitz et al., 1977; Phillips and Milo, 2009). The biological interpretation requires a Martian carbon-assimilation system whose relevant activity operates in an 8–26 °C incubation, survives the C5 pretreatment, including heating briefly at 120 °C and subsequently at 90 °C, and accounts for attenuation after combined storage and pretreatment at 175 °C. Horowitz et al. (1977) considered survival above 90 °C too robust for life expected on Mars. Newer data show, however, that the C5 temperatures are not necessarily sterilizing for all terrestrial organisms (Kempf et al., 2008; Schubert and Beaudet, 2011; Setlow and Setlow, 1995).

The Viking data do not conclusively establish such a microbial system on Mars, but they have not shown life to be inconsistent with the observations.

In an overview, although this analysis has focused on the CAPyR experiments, the results should also be considered in the context of Viking's full suite of life-detection experiments (Levin and Straat, 2016) (**Fig. 2**).

As a strength, that suite observed active chemical reactions, which (if they are biological) require an active life form. In contrast, much current discussion of biosignatures (National Academies of Sciences, Engineering, and Medicine, 2023) has returned to suggestions of biosignatures that statically analyze organic molecules, many of which are known to come from non-biological systems such as Bennu, Ryugu, and carbonaceous chondrites. The challenges of distinguishing biological organics from non-biological organics in such environments are well covered (Sephton et al., 2025).

Further, under their intended metabolic interpretations, LR and CAPyR probed oxidation and reduction in separate aliquots from common Martian soil acquisitions. Living systems require both oxidation and reduction reactions, often in a single pathway. Generally, both oxidation and reduction are required to break down input material (catabolism) to rearrange its resource atoms to build living material (anabolism).

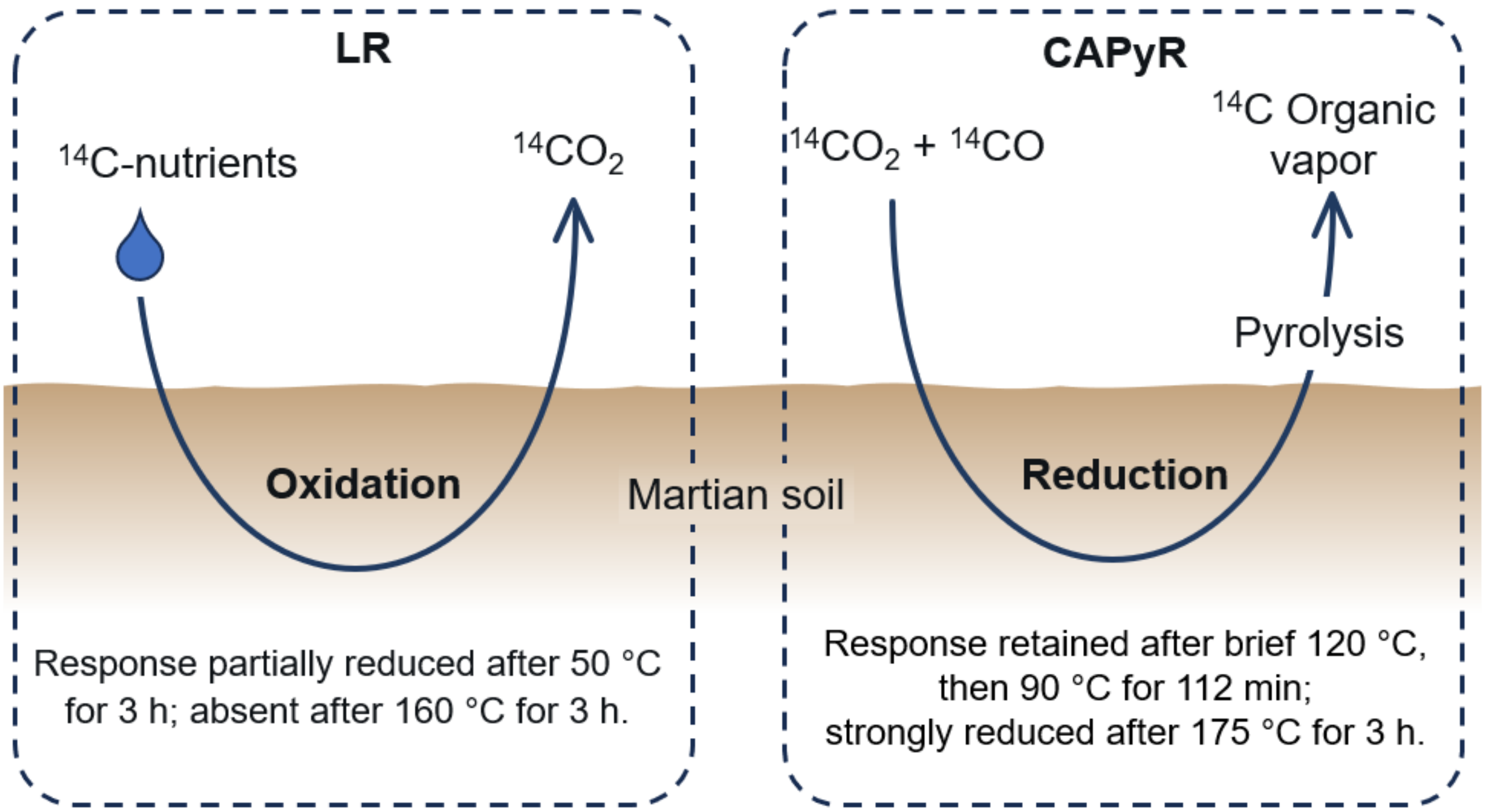

***Figure 2.*** *In living organisms, both oxidation and reduction processes are required to catabolize and anabolize organic materials to convert organic feedstocks into biomass. The Viking experiments probed (i) nutrient-derived radiocarbon release and (ii) retained radiocarbon incorporation in separate aliquots from common Martian soil acquisitions, interpreted in this conceptual diagram as oxidation and reductive assimilation, respectively.*

***(left)*** *According to the design, the Labeled Release (LR) experiment observed the oxidative conversion of supplied $^{14}$C-labeled organic nutrients to give approximately 15 nmol of $CO_2$-equivalent carbon per 0.5 $cm^3$ of soil, inferred using the $^{14}$C tracer (Klein, 1978). An oxidative interpretation requires an electron acceptor and sufficient reaction kinetics. In a biological model, that oxidant may be the mild $O_2$ oxidant that was released upon humidification in the gas exchange (GEx) experiment. Biological catalysts (enzymes) would let this kinetically slow oxidant produce the $^{14}CO_2$ at the rate* *that Viking observed. Any abiotic explanation involving $O_2$ or another oxidant must account for the observed kinetics under LR conditions, including possible mineral catalysis and the thermal and storage controls.*
***(right)** According to the design, the Carbon Assimilation–Pyrolytic Release (CAPyR) experiment observed the reduction of supplied $^{14}CO_2$ and/or $^{14}CO$ into material that remained after the 120 °C purge, but was released upon heating at 635 °C to give radiolabeled material that can be trapped as Peak 2, totaling 0.01 or 0.03 nmol total carbon per 0.25 $cm^3$ of soil in the C1 response, assuming derivation entirely from supplied CO or $CO_2$, respectively (Horowitz et al., 1977). Atmospheric $H_2$ or, with energy (photosynthesis), $H_2O$ could be biological reductants, again with biological catalysts (enzymes) that achieve the rates observed. However, no abiotic pathway examined here has been demonstrated to reproduce both the incorporation rate and the complete CAPyR treatment-response pattern. Any proposed pathway must also establish sufficient reductant availability, reaction time, product yield, and trap recovery.*

In Terran biology, oxidative and reductive processes coexist through the use of specific enzymes, cellular compartments, and separate cofactor pools. For example, the nicotinamide adenine dinucleotide ($NAD^+$) pool generally manages redox chemistry in catabolism, while the nicotinamide adenine dinucleotide phosphate ($NADP^+$) pool generally manages redox chemistry in anabolism.

The relevance to the Viking life detection suite is that explanations of the LR and CAPyR responses should be assessed together. However, heterogeneous regolith need not have a single equilibrium redox potential, and the assays used separate aliquots, feedstocks, and conditions. Both biological and abiotic models must therefore account quantitatively for the response classes and their contrasting treatment sensitivities.

The philosophical framework of the Viking team can be described as *eliminative induction*. They sought to rule out all other explanations before accepting biology. Klein said they had to be “forced into the biological explanation,” while Horowitz stated: “We hope by the end of this mission to have excluded all but one of the explanations, whichever that may be” (both quoted in Cooper, 1979a). The same philosophical framework is still applied today. No demonstrated combination of nonbiological processes explains the full pattern of observations examined here, while those observations remain consistent with biology. If this standard also requires excluding unknown chemical pathways, however, acceptance of biology can be delayed indefinitely as there will always be unknown abiotic pathways.

We suggest that *inference to the best explanation* is better suited to judge complex natural phenomena than *eliminative induction*. We propose to compare biological and nonbiological explanations by how well they explain the observations and withstand experimental tests, with the burden of explanation shifting as evidence accumulates. Accordingly, our reanalysis does not ‘prove’ life on Mars, but we argue that it shifts the burden of proof to proponents of a nonbiological origin. As a consequence, continued preference for abiotic interpretation requires evidence consistent with said abiotic interpretation. Referring to yet unknown chemistry remains a plausible source of uncertainty, but its mere possibility does not make an abiotic explanation more convincing.

Levin’s insistence that proposed oxidants be identified, quantified, and tested under Martian conditions remains a useful experimental criterion (Cooper, 1979a). Laboratory demonstrations, including perchlorate-radiolysis chemistry (Quinn et al., 2013), must be assessed for product inventories, reaction rates, and treatment responses under conditions relevant to Viking. The same requirement applies to agents proposed to account for CAPyR, including combinations of agents. In Cooper’s account, Lederberg asked what further instrumental observations could establish a biological origin (Cooper, 1979a). Future experiments should define such discriminating outcomes in advance and combine measurements of activity with independent molecular characterization. The unresolved Viking results justify these tests; they do not by themselves establish life on Mars.

Taken in overview, these considerations should redirect Mars exploration priorities back toward the direct search for extant life on Mars (Benner, 2026; Stoker et al., 2021), something that Jastrow understood already in 1977, but received less emphasis in subsequent exploration priorities, in part because of interpretations of the GC-MS results (Benner, 2026). This priority is also emphasized for the Martian subsurface by Nisson et al. (2026). Distributed, low-cost mission architectures such as IMPRESS provide one practical route for frequently conducted modern *in situ* life-detection experiments to multiple Martian surface and shallow-subsurface sites (Spacek et al., 2026).

## Authors' Contributions

All authors contributed to writing and editing the manuscript. Jan Spacek prepared the figures.

## Acknowledgments

We thank the CHiwi Foundation for supporting astrobiology research in our laboratory. We used ChatGPT-6 Astra (OpenAI) to assist with literature searches and manuscript phrasing refinement. Following the use of this tool, the authors independently reviewed, verified, and edited all generated content and references for factual accuracy. The authors retain full accountability and responsibility for the final content of the manuscript.

## Author Disclosure Statement

Jan Spacek is the founder and majority equity holder of IMPRESS Spaceworks Co. The authors declare no other competing interests.

## Funding Information

This work was supported by the CHiwi Foundation.